\documentclass[aps,prb,twocolumn,groupedaddress,amsmath,amssymb,amsfonts,graphics]{revtex4}

\newcommand{\bracket}[1]{\langle #1 \rangle}

\newcommand{\crit}{\text{c}}
\usepackage{graphicx}
\usepackage{bm}
\usepackage{bbold}
\renewcommand{\vec}[1]{\mathbf{#1}}

\begin{document}

\title{Valence bond description of the long-range, nonfrustrated Heisenberg chain}

\author{K.\ S.\ D.\ Beach}
\email[Email: ]{ksdb@physik.uni-wuerzburg.de}
\affiliation{Institut f\"{u}r Theoretische Physik, Universit\"{a}t W\"{u}rzburg, Am Hubland, 97074 W\"{u}rzburg, Germany}
\date{September 27, 2007}

\begin{abstract}
The Heisenberg chain with antiferromagnetic, powerlaw exchange
has a quantum phase transition separating spin liquid and N\'{e}el 
ordered phases at a critical value of the powerlaw exponent $\alpha$. 
The behaviour of the system can be explained rather simply in terms of a
resonating valence bond state in which the amplitude for a bond of 
length $r$ goes as $r^{-\alpha}$ for $\alpha < 1$,
as $r^{-(1+\alpha)/2}$ for $1 < \alpha < 3$, and as $r^{-2}$ for $\alpha > 3$.
Numerical evaluation of the staggered magnetic moment and Binder cumulant
reveals a second order transition at $\alpha_{\crit} = 2.18(5)$, in excellent agreement with
quantum Monte Carlo. The divergence of the magnetic correlation length is 
consistent with an exponent $\nu = 2/(3-\alpha_{\crit}) = 2.4(2)$.
\end{abstract}

\maketitle

\emph{Introduction}---Quantum spin-half chains whose interactions are local and only weakly 
frustrating\cite{Haldane82} have a quasi-long-range ordered ground state with powerlaw
spin correlations.~\cite{Affleck89a} This is different from
the situation in higher dimensions, where such models exhibit
true long-range order (LRO).~\cite{Sandvik97,Castro06}
It is well known that LRO in one dimension is proscribed
by theorem,~\cite{Bruno01} but only when the interactions are
sufficiently short-ranged.
With the addition of an antiferromagnetic interaction of 
arbitrary strength and range, the Heisenberg spin chain acquires a
phase diagram that includes both spin liquid and N\'{e}el-ordered regions.

Laflorencie and coworkers~\cite{Laflorencie05} have proposed a model of the form
$\hat{H} = \sum_{ij}J_{ij}\vec{S}_i\cdot\vec{S}_j$ with an
exchange coupling
\begin{equation} \label{EQ:Jij}
J_{ij} = \gamma_{ij} - \lambda(1-\gamma_{ij})\frac{(-1)^{i+j}}{\lvert r_{ij}\rvert^\alpha}.
\end{equation}
Here, $\lambda$ and $\alpha$ are positive parameters, $r_{ij}$ is the distance 
between sites $i$ and $j$, and 
$\gamma_{ij}=\delta(\lvert r_{ij}\rvert -1)$ is the nearest-neighbour (NN)
matrix. The authors of Ref.~\onlinecite{Laflorencie05} have mapped out the 
$\lambda$--$\alpha$ phase diagram using quantum Monte Carlo. They report the 
existence of a line of critical points---separating the magnetically 
disordered and ordered phases---along which the critical exponents vary 
continuously; the dynamical exponent obeys the inequality $z < 1$. 
Studies of the $\lambda=1$ model have previously
been carried out using a real-space renormalization
group method\cite{Rabin80} and spin wave theory.~\cite{Yusuf04} 

In this paper, we show that a valence bond (VB) description\cite{Rumer32} of the
long-range spin chain provides a unified picture of the quantum phase transition
and of the (seemingly) quite different ground states on either side of it. Moreover, we
show that numerical results based on a resonating valence bond (RVB) 
wavefunction are \emph{quantitatively} accurate, even near criticality.

As we have
argued elsewhere,~\cite{Beach07a,Beach07b} spin models with local, 
nonfrustrating interactions are well described by
RVB states~\cite{Liang88} in which bonds of length $r$ appear
with a probability amplitude $(\xi^2 + r^2)^{-(d+1)/2}$. Although this result
is most accurate for large dimension $d$, it remains a good 
approximation for spin-$S$ systems in $d=1$ provided that $2S$ is odd. 
The only subtlety is to explain why the properties of the $1/r^2$ state in 
$d=1$ and the $1/r^3$ state in $d=2$ are so different.

In the VB picture, the liquidness of the linear chain 
and the antiferromagnetism of the square lattice (for instance) are
consequences of their different lattice geometries. Overlaps of VB 
states form a collection of loops, and the existence of magnetic LRO 
is related to the loops being macroscopic in size.~\cite{Beach06}
In general, the relationship between bonds and loops is nontrivial.
For the family of RVB states with $r^{-p}$ bond amplitudes, it turns out
that there is a critical value of the exponent $p$ below which the typical 
loop becomes system-spanning (marking the onset of magnetic LRO). This 
critical value increases monotonically with the dimension of the lattice (and
diverges for $d=3$). For the linear chain the value is $p_{\text{c}} \approx 1.6$ 
and for the square lattice $p_{\text{c}} \approx 3.3$~\cite{Havilio99,Beach07a};
clearly, $d+1 > 1.6$ for $d=1$ (disorder), whereas $d+1 < 3.3$ for $d=2$ (order).

\begin{figure}
\includegraphics{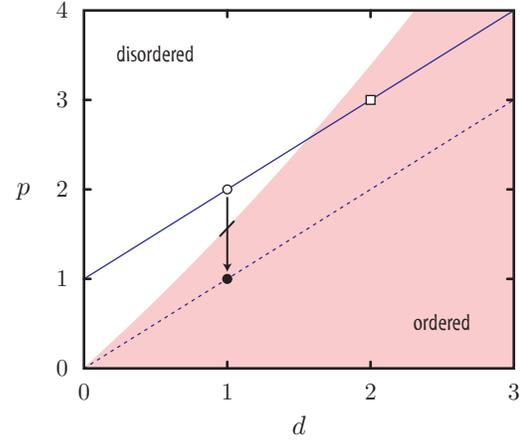}
\caption{\label{FIG:picture}
A slice in the space of RVB wavefunctions spanned by the lattice dimension $d$ 
and the powerlaw exponent $p$ of the long-bond tail. Spin 
models with local, nonfrustrating interactions live along the line $p = d+1$.
Those with long range, nonfrustrating interactions live in the band $d \le p \le d+1$.
The nearest neighbour Heisenberg model is indicated with an open circle ($d=1$)
and an open square ($d=2$). In one dimension, the path from $p=2$ to $p=1$ 
(from open circle to filled circle) passes through a magnetic transition.
}
\end{figure}

With the introduction of sufficiently long-range interactions (powerlaw 
decay with exponent $\alpha < 3$), the decay exponent of the bond amplitude 
function can be tuned continuously in the range $1 < p < 2$. See Fig.~\ref{FIG:picture}.
For the linear spin chain (in a model related to Eq.~\eqref{EQ:Jij} with $\lambda=1$), 
the critical $p_{\text{c}}$ is achieved at $\alpha_{\text{c}} = 2.18(5)$.
This compares favourably to the critical value $\alpha_{\text{c}}^{\text{qmc}} = 2.225(25)$
determined by quantum Monte Carlo in Ref.~\onlinecite{Laflorencie05}
and represents a significant improvement over the values
predicted by spin wave theory, $\alpha_{\text{c}}^{\text{sw}} = 2.46$,~\cite{Laflorencie05}
and by a numerical renormalization group method,
$\alpha_{\text{c}}^{\text{rg}} = 1.85$.~\cite{Rabin80}

\emph{RVB analysis}---The singlet ground state of an even number of $S=\tfrac{1}{2}$ spins can
be expressed in an overcomplete basis of bipartite VB states.~\cite{Beach06}
A simple RVB wavefunction can be constructed by assigning to each bond
of odd length $r$ an amplitude $h(r)$ and taking as the total weight
for each VB configuration the product of the
amplitudes of the individual bonds:
\begin{equation} \label{EQ:rvb}
\lvert h \rangle = \sum_{v} \biggl[ \prod h(r) \biggr] \lvert v \rangle.
\end{equation}
The sum in Eq.~\eqref{EQ:rvb} is over all possible pairings of spins
in opposite sublattices, 
and the product is over all bond lengths $r$ appearing in the VB
configuration.

\begin{figure}
\includegraphics{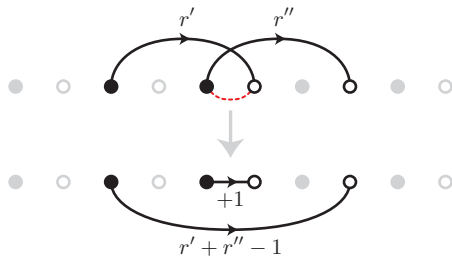}
\caption{\label{FIG:updates} A valence bond is a singlet constructed from
two spins in opposite sublattices and characterized by the directed distance
between them. In a valence bond state, all the spins are paired to form valence bonds in
one of the $(L/2)!$ possible configurations.
The reconfiguration induced by the nearest-neighbour Heisenberg interaction 
is sketched above. The dotted (red) line shows where the interaction is applied.
Bonds $r',r''$ are mapped to $1,r'+r''-1$.
}
\end{figure}

The proper choice of $h(r)$ is determined by the model at hand.
We observe that the application of a bipartite Heisenberg interaction to a VB state results
in the reconfiguation of bonds depicted in Fig.~\ref{FIG:updates}.
When acting between sites
$i$ and $j$ (in opposite sublattices), separated by a distance $a = r_{ij}$,
the interaction transforms bonds of length $r'$ and $r''$ 
into bonds of length $a$ and $r=r'+r''-a$.
The steady-state solution~\cite{Beach07b} of this reconfiguration process is
\begin{equation} \label{EQ:hq}
h(r) = \int_{-\pi/2}^{\pi/2}\!\frac{dq}{\pi}\,e^{{\mathrm i}qr}h_q,\ \ \  h_q = \frac{1-(1-J_q^2)^{1/2}}{J_q},
\end{equation}
where $J_q$ is the Fourier transform of the interaction
normalized to $J_{q=0}=1$. In a model with NN interactions only,
$J_q = \cos q$, and the long distance behaviour is
\begin{equation} \label{EQ:h1d}
h(r) = \frac{2}{\pi(1 + r^2)}.
\end{equation}

Numerical evaluation of Eq.~\eqref{EQ:rvb} can be
performed stochastically on lattices of finite size, as described in 
Ref.~\onlinecite{Beach07b}. We find that the RVB state for the Heisenberg chain
(characterized by Eq.~\eqref{EQ:hq} with $J_q = \cos q$) has a magnetically 
disordered ground state and an $L\to\infty$ extrapolated energy $E = -0.4360(1)$, 
within 1.7\% of the exact result $E = \log 2 - 1/4 = -0.44315$. The discrepancy 
is somewhat large, but not unreasonably so given that our wavefunction was not 
variationally determined. Moreover, the RVB state performs worst in the disordered 
phase; the agreement is increasingly good the deeper we go into the
magnetic region.

\begin{figure}
\includegraphics{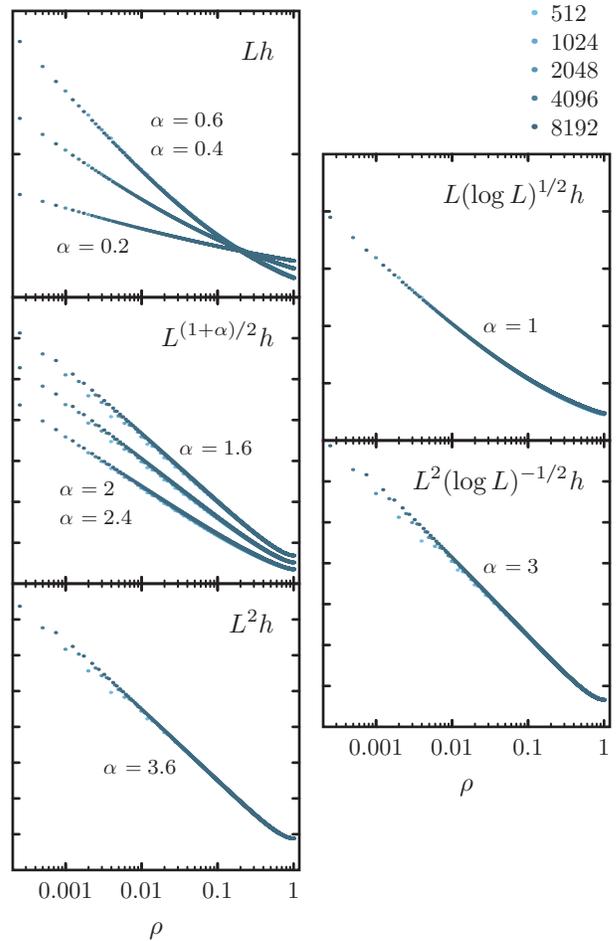}
\caption{ \label{FIG:collapse}
These log-log plots illustrate the five different scaling behaviours described
in Eq.~\eqref{EQ:scalingH}. Various renormalizations of the bond amplitude $h$
are plotted with respect to the reduced coordinate $\rho = 2r/L$.
All the vertical axes (labels suppressed) should be understood to start at $10^0$ 
and increase by one decade per tic ($10^1$, $10^2$, etc.).
}
\end{figure}

\begin{figure}
\includegraphics{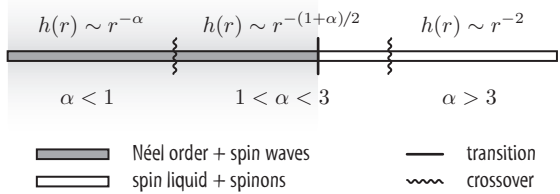}
\caption{\label{FIG:phases}
The long-bond behaviour of the amplitude function is powerlaw
irrespective of the range of interaction, but the decay exponent changes as
a function of $\alpha$. We identify three regimes. 
When $\alpha > 3$, the bond amplitudes share the $h(r) \sim r^{-2}$ 
and $\omega_q \sim q$ behaviour of the nearest-neighbour Heisenberg model, 
differing only in the amplitudes
of the short bonds and the value of the spinon velocity. 
In the intermediate regime $1 < \alpha < 3$, the decay exponent varies
continuously, and the spin waves exhibit sublinear 
dispersion, $\omega_q \sim q^{(\alpha - 1)/2}$. 
Below a critical ($\lambda$-dependent) $\alpha_{\crit}$, 
the spins develop long-range antiferromagnetic order. When $\alpha < 1$, the
interaction is superextensive and the system is close to being classically N\'{e}el
ordered. 
}
\end{figure}

We now consider the long-range exchange integral
\begin{equation}
J_{ij} \sim \frac{1-(-1)^{i+j}}{\lvert r_{ij} \rvert^{\alpha}},
\end{equation}
which is equivalent to the $\lambda=1$ case in Eq.~\eqref{EQ:Jij},
except that we have removed the interactions between spins in the same sublattice
and compensated with coupling strength of opposite sign at neighbouring sites.
This change has no real significance but it does simplify our analysis since
nonbipartite interactions would require introducing a second update rule
(different from the one shown in Fig.~\ref{FIG:updates}).

In this case, the $J_q$ to appear in Eq.~\eqref{EQ:hq} is 
\begin{equation} \label{EQ:Jqlong}
J_q = 1+ \frac{\sum_n [\cos(nq)-1] n^{-\alpha}}{\sum_n n^{-\alpha}}
= 1 - \tfrac{1}{2} \xi^2_\alpha q^2 + \cdots 
\end{equation}
Here,  $n$ runs over all odd integers from $1$ to $L/2-1$,
and, to leading order in $L$,
\begin{equation}
\xi_{\alpha}^2  = \frac{\sum_n n^{2-\alpha}}{\sum_n n^{-\alpha}}
= \begin{cases}
\frac{1}{4}(\frac{1-\alpha}{3-\alpha})L^2 & \alpha < 1\\
\frac{L^2}{8\log L} & \alpha=1 \\
\frac{L^{3-\alpha}}{(3-\alpha)2^{4-\alpha}(1-2^{-\alpha})\zeta(\alpha)} & 1 < \alpha < 3\\
\frac{4 \log L}{7\zeta(3)} & \alpha = 3 \\
\frac{(1-2^{2-\alpha})\zeta(\alpha-2)}{(1-2^{-\alpha})\zeta(\alpha)} & \alpha > 3
\end{cases}
\end{equation}
When $\alpha > 3$, $\xi_\alpha$ is O(1), and the long wavelength 
behaviour of 
$h_q = e^{-\xi_\alpha q} + O(\xi_{\alpha}^3q^3)$ is integrable; hence,
\begin{equation} \label{EQ:halphagt3}
h(r) = \frac{2\xi_{\alpha}}{\pi(\xi_{\alpha}^2+r^2)}.
\end{equation}
This is the same form we found for $J_q = \cos q$,
except that $\xi_\alpha$, which controls the length scale
at which the $r^{-2}$ tail is cut off, is now $\alpha$ dependent.
In the limit $\alpha \to \infty$, $\xi_\alpha \to 1$, and we recover Eq.~\eqref{EQ:h1d}.

As $\alpha \to 3^+$,  $\xi_{\alpha}$ diverges, and $h(r)$ changes functional form
to $r^{-2}(\log r)^{1/2}$ at $\alpha = 3$. There is an additional crossover at $\alpha = 1$, where
$h(r)$ goes as $r^{-1}(\log r)^{-1/2}$. Otherwise, the bond amplitude has
a continuously variable decay exponent: $r^{-\alpha}$ for $\alpha < 1$, and $r^{-(1+\alpha)/2}$
for $1 < \alpha < 3$. These results can be demonstrated by noting
that $h(r)$ obeys the relation
\begin{equation}
h(r) = \frac{\xi_{\alpha}}{L^2} H_\alpha\biggl(\frac{2r}{L}\biggr)
\end{equation}
with a scaling function
\begin{equation} \label{EQ:scalingH}
H_\alpha(\rho) \sim \begin{cases}
\rho^{-\alpha} & \alpha < 1 \\
\rho^{-1}(\log \rho)^{-1/2} & \alpha = 1\\
\rho^{-(\alpha+1)/2} & 1 < \alpha < 3 \\
\rho^{-2}(\log\rho)^{1/2} & \alpha = 3 \\
\rho^{-2} & \alpha > 3
\end{cases}
\end{equation}
as verified by data collapse in Fig.~\ref{FIG:collapse}.

\begin{figure}
\includegraphics{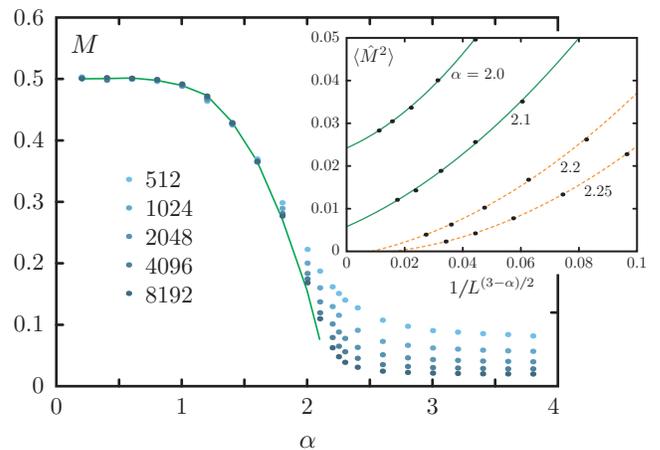}
\caption{\label{FIG:mag1d}
$\bracket{\hat{M}^2} = (1/L^2)\sum_{ij}(-1)^{i+j}\bracket{\vec{S}_i\cdot\vec{S}_j}$,
the staggered structure factor, is computed for lattices up $L=8192$ in size.
The main plot shows the magnitude of
the staggered magnetic moment, $M = \bracket{\hat{M}^2}^{1/2}$, as a function
of $\alpha$. The value of $M$ in the thermodynamic limit is marked with
a solid (green) line.
The inset shows the finite-size scaling of $\bracket{\hat{M}^2}$ in the
vicinity of the magnetic quantum critical point. The extrapolations to $1/L= 0$
are draw with solid (green) or dashed (orange) lines depending on 
whether the intersection with the y-axis is positive. We estimate that
the critical point lies in the range $2.1 < \alpha_{\crit} < 2.2$.
}
\end{figure}

From what we know about the RVB ground state, it is possible to make a
reasonable guess for the excitation spectrum $\omega_q$. Wherever the system 
is N\'{e}el ordered, the low-lying excitations are magnons. These spin-1
excitations can be created by promoting in succession each singlet
bond in the RVB state to a triplet (along with the appropriate $q$-dependent
phase factor): at the mean field level, one finds that
$h_q \sim e^{-\omega_q}$ in the $q\to 0$ limit.~\cite{Beach07a}
With this identification, $\omega_q\sim q$ when $\alpha > 3$, and 
$\omega_q\sim q^{(\alpha-1)/2}$ when $1 < \alpha < 3$.

Numerical evaluation of the RVB wavefunction over a range of $\alpha$
reveals the phase diagram summarized in Fig.~\ref{FIG:phases}. We find that
the staggered moment, shown in Fig.~\ref{FIG:mag1d}, is nearly saturated 
at $M=0.5$ throughout the $\alpha < 1$ semi-classical region. $M$ decreases monotonically
across the intermediate regime $1 < \alpha < 2$ and vanishes continuously at
some $\alpha = \alpha_{\crit}$. Quantum disorder for $\alpha > 3$ is
guaranteed by theorem.~\cite{Parreira97}

\begin{figure}
\includegraphics{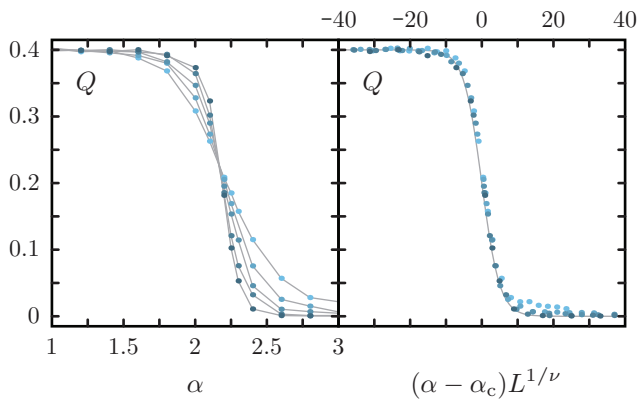}
\caption{\label{FIG:binder}
(Left) $Q = 1 - \bracket{\hat{M}_z^4}/3\bracket{\hat{M}_z^2}^2$,
the Binder cumulant, is plotted as a function of
$\alpha$ for lattices of increasing size. (The curves have increasing slope,
approaching a step function for $L\to\infty$). The data points
are shaded as in Figs.~\ref{FIG:collapse} and \ref{FIG:mag1d}.
The lines, connecting points of equal $L$, emphasize the invariance
of $Q$ at $\alpha=\alpha_{\crit}$. (Right) The same data are replotted
with an $\alpha_{\crit}$ shift and $L^{1/\nu}$ rescaling of the
horizontal axis. Good data collapse is acheived for $\alpha_{\crit} = 2.18$
and $\nu = 2/(3-\alpha_{\crit}) = 2.44$. The line
$(1/5)[1-\tanh(x/5)]$ is drawn as a guide to the eye.
}
\end{figure}

Under the assumption of $\omega_q\sim q^{(\alpha-1)/2}$ spin excitations,
the staggered structure factor exhibits $1/L^{(3-\alpha)/2}$ 
scaling.~\cite{Laflorencie05} Accordingly, the inset in Fig.~\ref{FIG:mag1d}
shows it vanishing somewhere in the range $2.1 < \alpha_{\crit} < 2.2$. 
A more sophisticated finite-size scaling
analysis (see Fig.~\ref{FIG:binder}), 
based on data collapse of the Binder cumulant,~\cite{Binder81}
indicates $\alpha_{\crit} = 2.18(5)$. The correlation length exponent extracted
from the fit, $\nu = 2.4(2)$, is in agreement with 
the large-$N$ prediction of the corresponding $N$-component vector theory:~\cite{Laflorencie05}
namely, $\nu = 1/(\alpha_{\crit}-1)$ for $1 < \alpha_{\crit} \le 5/3$, and
$\nu = 2/(3-\alpha_{\crit})$ for $5/3 \le \alpha_{\crit} < 3$.

The transition is in the quantum percolation class.~\cite{Vojta05}
As the range of interaction is ramped up, the bonds in the RVB state grow longer,
and the overlap loops increase in size. Across the magnetic transition, the scaling dimension 
$D$ of the average loop size ($\sim L^D$) changes discontinuously from $D=0$ to $D=1$. 
At criticality, the loops have fractal dimension $0 < D_f < 1$, and the dynamical exponent is
fixed by $z = D_f$. This explains the observation in Ref.~\onlinecite{Laflorencie05} 
that $z<1$. (Analogous behaviour~\cite{Beach07a} is found in two dimensions for radially 
symmetric bond amplitude functions.)

\emph{Conclusions}---The NN Heisenberg chain has a quantum disordered 
ground state, but the addition of long-range, antiferromagnetic interactions 
can drive the formation of magnetic LRO. The ground state wavefunction evolves
smoothly across the phase boundary. Its structure on both sides of the transition 
is essentially that of a factorizable RVB wavefunction whose bond amplitudes 
decay as a powerlaw in the bond length. Only the value of the decay exponent 
changes across the transition.

The onset of N\'{e}el order can be understood as a quantum percolation transition
in which the VB loops become system-spanning. The dynamical exponent 
$0 < z < 1$ is equal to the fractal dimension of the VB loops that are formed 
at criticality. The exact point at which the transition occurs and the value of the 
critical exponents there depend sensitively on the bond distribution.

The RVB picture is often invoked in a loose, heuristic way. Here, we have
presented a concrete RVB wavefunction that proves to be a practical 
computational tool. Many aspects of the unbiased quantum 
Monte Carlo results can be reproduced by the RVB wavefunction
with dramatically less computational effort.

The author acknowledges helpful discussions with Nicolas Laflorencie.
This work was supported by the Alexander von Humboldt Foundation.

\end{document}